
\input phyzzx

\overfullrule=0pt
\font\twelvebf=cmbx12
\nopagenumbers
\footline={\ifnum\pageno>1\hfil\folio\hfil\else\hfil\fi}
\line{\hfil CU-TP-611}
\line{\hfil IASSNS-HEP-93/33}
\line{\hfil NSF-ITP-93-135}

\vglue .4in
\centerline {\twelvebf  Vortex Dynamics in Selfdual Maxwell-Higgs Systems }
\centerline{\twelvebf  with  Uniform  Background Electric
Charge Density$^\dagger$}
\vskip .5in
\centerline{\it  Ki-Myeong Lee }

\vskip .1in
\centerline {$^*$Physics Department, Columbia University}
\centerline {New York, New York 10027}
\vskip .15in
\centerline{and}
\vskip .15in

\centerline{Institute for Advanced Studies}
\centerline{Olden Lane, Princeton, NJ 08540}
\vskip .15in
\centerline{and}
\vskip .15in

\centerline{Institute for Theoretical Physics}
\centerline{ University of California at Santa Barbara}
\centerline{ Santa Babara, CA 93106}

\baselineskip=18pt
\overfullrule=0pt

\vskip .5in
\centerline {\bf Abstract}
\vskip .1in
We introduce selfdual Maxwell-Higgs systems with uniform background
electric charge density and show that the selfdual equations satisfied
by topological vortices can be reduced to the original Bogomol'nyi
equations without any background.  These vortices are shown to  carry no spin
but to feel the Magnus force due to the shielding charge carried by the
Higgs field.  We also study the dynamics of slowly moving vortices and
show that the spin-statistics theorem holds to our vortices.

\vfill

\noindent November 1993
\footnote{}{$^\dagger$  This work is supported in part
by the NSF Presidential Young Fellowship, A. Sloan Fellowship,  NSF Grant
PHY-92-45317 and NSF Grant PHY-89-04035 }

\footnote{}{$^*$ Permanent Address. Email Address:
klee$@$cuphyf.phys.columbia.edu}
\vfill\eject

\def\pr#1#2#3{Phys. Rev. {\bf D#1}, #2 (19#3)}
\def\prl#1#2#3{Phys. Rev. Lett. {\bf #1}, #2 (19#3)}

\def\np#1#2#3{Nucl. Phys. {\bf B#1}, #2 (19#3)}
\def\pl#1#2#3{Phys. Lett. {\bf #1B}, #2 (19#3)}

\REF\rNielsen{ A.A. Abrikosov, Sov. Phys. JETP {\bf 5},
    1174 (1957); H.B. Nielsen and P. Olesen, \np{61}{45}{73}.}

\REF\rDavis{  R.L. Davis,
Mod. Phys. Lett. {\bf A5}, 853 (1990). R.L. Davis, Mod. Phys. Lett.
{\bf A5}, 955 (1990) }

\REF\rBarcall{S. Bahcall and L. Susskind, Int. Mod. Phys. B,  2735 (1991).}

\REF\rBogo{E.B. Bogomol'nyi, Sov. J. Nucl. Phys. {\bf 24}, 449 (1976).}

\REF\rAo{ P. Ao and D.J. Thouless, \prl{70}{2158}{93}.}

\REF\rSonin{ E.B. Sonin, Rev. Mod. Phys. {\bf 59}, 87 (1987). }

\REF\rKim{ Y.B. Kim and K. Lee, `Vortex Dynamics in Selfdual
Chern-Simons-Higgs Systems,' Columbia and CERN preprint, CU-TP-574 and
CERN-TH.6701/92.}

\REF\rManton{N.S. Manton,\pl{110}{54}{82}.}

\REF\rRuback{  R. Ruback. \np{296}{669}{88};
K.J.M. Moriaty, E. Meyers and C. Rebbi, \pl{207}{411}{88}; E.P.S.
Shellard and P.J. Ruback, \pl{209}{262}{88};
T.M. Samols, Comm. Math. Phys. {\bf 145}, 149 (1992); E. Myers, C. Rebbi
and R. Strika, \pr{45}{1355}{92}.}

\REF\rWeinberg{E.J. Weinberg, \pr{20}{3008}{79}; C.H. Taubes, Commun.
Math. Phys. {\bf 72}, 277 (1980).}

\chapter{Introduction}

We consider the theory of complex and neutral scalar fields coupled to
a gauge field with the Maxwell kinetic term in three dimensions. There
exist topological vortices of nonzero magnetic flux in the broken
phase.\refmark\rNielsen We introduce a nondynamical uniform background
electric charge density to the system, which would be shielded by the
electric charge carried by the Higgs field. In this case, vortices are
claimed to carry nonzero spin and feels the fluid dynamical Magnus
force, a Lorentz force, due to the shielding charge carried by the
Higgs field.\refmark\rDavis In addition, vortices in charged fluid
with a nondynamical background magnetic field have been studied to see
whether they are anyons.\refmark\rBarcall Here we choose a set of special
values for coupling constants so that there is a bound on energy,
which can be saturated by the vortex configurations satisfying the selfdual
equations, generalizing the Bogomol'nyi case.\refmark\rBogo We study
in detail the vortex dynamics in this selfdual model.

Recently, there has been a renewed interest in the possibility of the
Magnus force in real superconductor.\refmark\rAo The question is
whether there is a nonzero Berry's phase gained by a vortex wave
function due to the Magnus force when it goes around a closed loop on
a plane. In real superconductors, the copper pairs condense with net
nonzero electric charge, which is neuralized by the background
electrons and positive ions. The Maxwell-Higgs system with the
background electric charge density is thus more closely related to the
real superconductor rather than the Maxwell-Higgs system without as
noted by Davis.\refmark\rDavis Our analysis confirms the presence of the
Magnus force more clearly. Since our system is selfdual or there is no
static interaction between vortices, we can regard our model lying at
the boundary between type I and type II superconductors.

The Magnus force is the fluid dynamical force responsible for curve
balls. The Magnus force also plays an important role in the vortex
dynamics in superfluid.\refmark\rSonin In superfluid, vortices carry
infinite angular momentum density per unit length and feel finite
amount force per unit length.  In large distance, the Magnus force can
induce a Berry's phase.  For example, in Chern-Simons-Higgs systems,
vortices carry nonzero magnetic flux and charge.  These vortices carry
nonzero spin and their statistics can be explained only when one put
together the naive Aharanov-Bohm phase due to the charge and magnetic
flux and that due to the Magnus force.\refmark\rKim

In our case, it turns out to be subtle to define the conserved angular
momentum in the field theory due to the divergent contribution from
the spatial infinity. We provide a satisfactory modification of the
Noether angular momentum in the field theory. The selfdual
configurations are degenerated in energy but not in angular momentum.
The total angular momentum is a complicated function of the vortex
positions. We show that there is however no intrinsic spin carried by
our vortices, which does not contradict the fact that our vortices
carry nonzero magnetic flux but no electric charge. In our systems,
the magnus force is in a way decoupled from the spin of vortices.

As there are no massless excitations in our systems as we will see, we
expect little or no radiations emitted when vortices are moving very
slowly. The field configuration of these slowly moving vortices
would be very close to that of vortices at rest. There would
be an effective Newtonian action describing these vortices, which
would be made of terms linear and quadratic in velocities. The
interesting goal would be then to find this effective action.  The
effective action for slowly moving solitons has been first studied by
Manton.\refmark\rManton There have numerical and analytical studies of
this effective action for slowly moving vortices when there is no
background charge.\refmark\rRuback

As these approaches are not directly applicable in our case, we take a
little bit different approach.\refmark\rKim We find the first order
terms explicitly. The angular momentum calculated from this effective
action turns out to be identical to that obtained from the field
theory. The linear term has an interesting implication in the vortex
dynamics. A single vortex would move a circle due to the Magnus force,
which implies a nontrivial Berry's phase in quantum mechanics. For a
system of two vortices, the Magnus force become more complicated when
two vortices are close to each other since the charge density is not
uniform. When two vortices are separated in large distance, we will
show that however there is no additional Berry's phase which can be
attributed to the statistics between vortices, confirming the
spin-statistics theorem for our vortices.

This paper is organized as follows.  In Sec.~2 we introduce the
selfdual Maxwell-Higgs systems with uniform background electric charge
density and study their basic properties. We show that the naive
conserved linear and angular momenta have divergent contributions from
the spatial infinity. We show that our selfdual equations can be
reduced to those found by Bogomol'nyi.  In Sec.~3, we study the
rotationally symmetric vortices numerically. In Sec.~4 we redefine the
linear and angular momenta  and show that our vortices do not carry
any spin. In addition, we provide an explicit expression of the
angular momentum as a function of vortex positions.  In Sec.~5, we
study the effective action of slowly moving vortices. This action
contains the terms linear and quadratic in vortex velocities. The linear
terms describe the magnetic interaction between vortices themselves
and between vortices and the shielding charge carried by the Higgs
field.  We use this linear term to calculate the statistics of our
vortices.  In Sec.~6, we conclude with some remarks and questions. In
the appendix, we calculate the nonconserved angular momentum derived
from the symmetric energy momentum tensor as a function of vortex
positions.

\chapter{Model}

We consider the theory of charged and neutral scalar fields $\phi=
fe^{i\theta} /\sqrt{2}, N$ coupled to a photon field $A_\mu$. We
assume that there is a unform background electric charge density
$\rho_{\rm e}$, which is no dynamical.  (A uniform external magnetic
field plays a role of chemical potential for a magnetic flux after a
field shift and so no role in the classical dynamics of vortices.)
The lagrangian for this theory is
$$ \eqalign{ {\cal L} = &\ -{1\over 4} F_{\mu\nu}^2 + {1\over
2}(\partial_\mu N)^2 + {1\over 2}
(\partial_\mu f)^2 \cr
&\ + {1\over 2} f^2 (\partial_\mu \theta + e A_\mu )^2 -
U(f,N,  \rho_{\rm e}) -
 \rho_{\rm e} A_0   \cr}
\eqno\eq $$
The Lorentz symmetry is explicitly broken due to the external electric
charge density.  The charge conjugation, $\theta \rightarrow -\theta,
\,\, A_\mu \rightarrow -A_\mu$, is also explicitly broken by the
external charge density. The parity transformation, $(x^1,x^2) \rightarrow
(x^1,-x^2), (A_1, A_2 ) \rightarrow  (A_1,-A_2)$, is not broken.
The time reversal, $t \rightarrow -t, A_0 \rightarrow -A_0$, is also
explicitly broken. However, CTP is still a good symmetry.
Usually the selfdual systems are related to the $N=2$ supersymmetry with
a central term,  and the neutral scalar field and the gauge field are
a part of the vector multiplet.

The system is invariant under the local gauge transformation,
$\theta \rightarrow
\theta + \Lambda, A_\mu \rightarrow A_\mu - \partial_\mu \Lambda /e$, leading
to
Gauss's constraint from the variation of $A_0$,
$$ \partial_i F_{0i} + e f^2 (\dot{\theta} + eA_0 ) - \rho_{\rm e}
= 0  \eqno\eq $$
where the dot denotes the time derivative. Since the background charge
density is uniform, the  action is invariant
under the spacetime translation and there is a  conserved energy
momentum tensor  as a Noether current,
$$ T_{\mu\nu} = -F_{\mu\rho} \partial_\nu A^\rho + \partial_\mu N
\partial_\nu N + \partial_\mu f \partial_\nu f +
f^2(\partial_\mu \theta + eA_\mu )\partial_\nu \theta - \eta_{\mu\nu}
{\cal L} \eqno\eq $$
satisfying, $\partial^\mu T_{\mu\nu} = 0 $. The total energy $E = \int
d^2r T_{00} $ can be expressed as
$$ E = \int d^2 x {\cal E} \eqno\eq $$
after a partial integration and using Gauss's law (2.2),
where  the energy density is given by
$$ \eqalign{
{\cal E} = &\ {1 \over 2 } ( F_{0i}^2 + F_{12}^2) + {1\over 2} [
\dot{N}^2 + (\partial_i N)^2]
 + {1 \over 2} [\dot{f}^2 + (\partial_i f)^2] \cr
&\ + { 1\over 2} f^2 (\dot{\theta} + e A_0 )^2 + { 1\over 2} f^2 (\partial_i
\theta + e A_i )^2
 + U   \cr}
\eqno\eq $$

The conserved linear momentum $\int d^2x T^{0i}$ is
$$ \tilde{P}^i = -\int d^2x \left\{ F_{0j} \partial_i A_j + \dot{N}\partial_i
N + \dot{f}\partial_i f + f^2 (\dot{\theta} + eA_0 ) \partial_i \theta
\right\} \eqno\eq $$
Under the local gauge transformation $\theta
\rightarrow \theta + e\Lambda $ and $A_i \rightarrow A_i - \partial_i
\Lambda$, $T_{0i}$ is invariant up to a total derivative due to
Gauss's constraint, leaving the linear momentum invariant.
As the action is also invariant under rotation, there is a conserved
angular momentum current,
$$ {\cal J}_\mu = - \epsilon_{ij} F_{\mu i} A_j - \epsilon_{ij} x^i
T_{\mu j} \eqno\eq $$
with  $T_{\mu\nu}$  given by Eq.(2.3), satisfying $\partial^\mu {\cal
J}_\mu = 0 $.
The conserved Noether angular momentum $\tilde{J} = \int d^2 r {\cal J}_0  $
$$ \tilde{J} = -\int d^2x
\left\{ \epsilon_{ij} F_{0i}A_j + \epsilon_{ij}x^i T_{0j} \right\}
\eqno\eq $$
which is also invariant under local gauge transformations.
Note that under the translation of the whole system by $x^i
\rightarrow x^i + a^i $, $J \rightarrow J + \epsilon_{ij} a^i P^j $.

It turns out that the above definition of the momenta is not entirely
correct.  When there are $n$ vortices, we will see that $\partial_i
\theta \rightarrow -{n \over r} \hat{\varphi}^i $ and $ef^2 A_0
\rightarrow {\rho_e \over e} $
for large $r$. Thus, there is a divergent contribution from the
spatial infinity to these momenta. The definition of finite,
well-defined and conserved momenta will be given in Sec.4.

For the reasons which will be clear in a moment, we consider a
specific potential

$$ U = {e^2 \over 2} N^2 f^2 + {e^2 \over 8} (f^2 - v^2 )^2
-   \rho_{\rm e}  N
\eqno\eq$$
This potential has a term for the interaction between the external
charge and the neutral scalar field. While  the physical origin
of this interaction is  obscure, we can imagine such possibility.
This potential is not bounded from below, which we will see is not a
problem due to Gauss's law.

After some algebra and by using Gauss's law~(2.2),
we get for any physical configuration
$$ \eqalign{ {\cal E} = &\ {1 \over 2}(F_{0i}  +  \partial_i N)^2
+ {1\over 2} [F_{12} \pm
	 {e \over 2} (f^2 - v^2)]^2  +  {1\over 2} \dot{N}^2   \cr
&\  +{1\over 2} \dot{f}^2 + {1\over 2} f^2 [\dot{\theta} + e A_0
 -  eN ]^2
   + {1\over 2} [\partial_i f \pm \epsilon_{ij}f(
\partial_j \theta + e A_j  )]^2 \cr
&\ \pm  { ev^2 \over 2} F_{12}
 -  \partial_i [ F_{0i} N + {1 \over 2} \epsilon_{ij}
f^2(\partial_j \theta + e A_j )] \cr}
\eqno\eq $$
The ground state of this energy functional will be in phase where
$<f^2> \ne 0$.  As  there will be no massless mode, the gauge invariant
quantities, $F_{0i}, \partial_i \theta + eA_i $ would vanish
exponentially at the spatial infinity, making the boundary contribution
vanish. After integrating Eq.(2.10) over space, we get a bound on the total
energy,
$$ E \ge \pm  {e v^2 \over 2} \int d^2 r F_{12}    \eqno\eq $$
because the rest of terms are positive definite.

This  bound is saturated by the time independent configurations in a gauge
$\dot{\theta} = 0$ which satisfy  Gauss's law (2.2)  and
$$  \eqalign{ &\   A_0 - N =0  \cr
 &\  F_{12} \pm  {e \over 2} (f^2 - v^2)  = 0 \cr
 &\ \partial_i f \pm \epsilon_{ij} f(\partial_j \theta + eA_j ) = 0
\cr }
\eqno\eq $$
For these  selfdual configurations Gauss's law (2.2) becomes
$$ \partial_i^2 A_0 + \rho_e - e^2 f^2 A_0 = 0  \eqno\eq $$
Note that Eq.(2.12) for $A_i, f,\theta$ is identical to the selfdual
vortex equations obtained by Bogomol'nyi.\refmark\rBogo As far as the
scalar field and the vector potential are concerned, the selfdual
vortex configurations without background charge density is identical
to those with. The existence and uniqueness of the selfdual solutions
for the scalar field and the vector potential, describing the $2n$
parametered configurations of $n$ vortices, have been studied and
proven.\refmark\rWeinberg We will see that $A_0$ satisfying Eq.(2.13)
can be expressed explicitly in terms $f$. Thus, in our system also
there exist unique selfdual configurations of vortices parameterized
by the vortex positions.

The ground state is a homogeneous configuration of zero energy. From
Eqs.~(2.12) and (2.13) we get this ground state described  classically by
$$ \eqalign{ &\  f = v \cr
             &\ N = A_0 = {\rho_e \over e^2v^2} \cr
             &\ \theta = A_i = 0  \cr}
\eqno\eq $$
in a unitary gauge. (We can imagine inhomogeneous configurations,
which have a finite region where $f=0$ and $N$ being arbitrary large,
making the contribution from the potential energy to be arbitrary
negative.  The bound (2.11) tells us that these configurations have
positive energies.)  The small fluctuation analysis around this vacuum
implies that in the unitary gauge
$$\eqalign{ &\ ( w^2 - \vec{k}^2 -e^2v^2 ) \delta f + {2\rho_e \over v}
(\delta A_0 -  \delta N) = 0 \cr
&\ (w^2 -\vec{k}^2 - e^2 v^2 ) \delta N - {2\rho_e \over v} \delta f = 0 \cr
&\ (w^2 - \vec{k}^2 - e^2v^2 ) \delta \vec{A} + \vec{k} \vec{k}\cdot \delta
\vec{A} -  w \vec{k} \delta A_0 = 0 \cr
&\ (\vec{k}^2 + e^2v^2) \delta A_0 - w\vec{k} \cdot \delta \vec{A} +
{2\rho_e \over v} \delta f = 0 \cr}
\eqno\eq $$
where $\delta f, \delta N, \delta A_\mu \sim e^{iwt + i\vec{k}\cdot
\vec{r}} $.  There are three different  massive modes in long distance, or
small $\vec{k}$, with eigenvalues,
$$ \eqalign{ &\ w^2 = e^2 v^2 \cr
          &\ w^2 = e^2v^2 \left\{ 1 + 2 {\rho_e^2 \over e^4v^6}
[ 1 \pm \sqrt{ 1 + { e^4v^6 \over \rho_e^2} } ]  \right\}
\cr}
\eqno\eq $$
There is no instability due to these modes. We can see easily that the
first spectrum describes a vector boson of spin $\pm 1$ and the last two
spectra
describe two scalar bosons.

For the selfdual configurations of the positive magnetic flux, we
choose the upper sign of Eq.(2.12),
$$  \eqalign{ &\  F_{12}  + {e \over 2} ( f^2 - v^2 ) =0 \cr
    &\ \partial_i \theta + eA_i -  \epsilon_{ij} \partial_j \ln f  =0
\cr}
\eqno\eq $$
For the positive magnetic  flux configurations, the vorticity in $\theta$ turns
out negative as we will see.  We describe  $n$ vortices located at
positions $\vec{q}_a$  by
$$ \theta (\vec{r})  = - \sum_{a=1}^n  Arg( \vec{r} - \vec{q}_a ) \eqno\eq $$
satisfying
$$
\epsilon_{ij} \partial_i \partial_j \theta
 = -\partial_i^2 \sum_a \ln |\vec{r}-\vec{q}_a| =
-2\pi \sum_a \delta(\vec{r} -\vec{q}_a)
\eqno\eq $$
We see that $f$ should vanish as $|\vec{r} -\vec{q}_a|$ at
the position of each vortex for the complex scalar field $\phi$ to
behave well. Putting together Eqs.(2.17) and (2.19),  we get
$$ \partial_i^2 \ln f^2 - e^2(f^2 - v^2  ) = 4\pi \sum_a
\delta(\vec{r} - \vec{q}_a)   \eqno\eq $$
In addition, we see that the excessive flux
$\Psi = \int d^2 x F_{12}  =  \oint_\infty d\vec{s}\cdot \vec{A} = - {1\over e}
\oint_\infty d\vec{s} \cdot \vec{\nabla} \theta = 2\pi n/e$ is positive.

For a given $f$ configuration satisfying Eq.(2.20), Gauss's law (2.13)
determines the $A_0$ configuration. It turns out that $A_0$ can be solved
explicitly in terms of $f$,
$$ \eqalign{ A_0(\vec{r};\vec{q}) &\ = {\rho_e \over e^2 v^2}\left\{ 1
-  \sum_{a=1}^n (\vec{r}-\vec{q}_a)
\cdot {\partial \over \partial \vec{q}_a }
\ln f  \right\} \cr
&\ =  {\rho_e \over e^2 v^2} \left\{ 1+ n - \sum_a (\vec{r} -
\vec{q}_a)\cdot {\partial \over \partial \vec{q}_a} \ln {f \over
\prod_b |\vec{r} -\vec{q}_b| } \right\}  \cr}
 \eqno\eq $$
Notice  that the quantity in the right side is regular at
the vortex positions and approachs the right asymptotic value at spactial
infinity  and that $\partial/ \partial \vec{r} = - \sum_a \partial
/\partial \vec{q}_a$  on $f$ due to the space translation invariance.

\chapter{Rotationally symmetric solutions}

The rotationally symmetric ansatz for $n$ positive magnetic flux
vortices at origin is given in the rotational coordinate $(r,\varphi)$
as $f(r), \theta = -n\varphi, A_\varphi(r), A_0(r)$. The total
magnetic flux is then  $2\pi n/e$.
Eq.(2.17)  become
$$ \eqalign{ &\ e A_\varphi = n - r {d\ln f \over  dr }  \cr
&\ {1 \over r} {dA_\varphi \over dr} = {e \over 2} (v^2 - f^2) \cr
\cr} \eqno\eq $$
We introduce dimensionless quantities, $evr = s, f^2= v^2 F $,
Eq.(3.1) can be put as
$$  [s (\ln{ F \over s^{2n}} )']' + s(1-F) =  0
\eqno\eq $$
with the prime denotes $d/ds$.  The boundary conditions  are
$F(s)\sim s^{2n}$ for small $s$ and $F(\infty) =1$.

The scalar potential in Eq.(2.21) become
$$ A_0(r) = {\rho_e \over e^2v^2} \left\{
1 + {s \over 2} (\ln F)' \right\}
\eqno\eq $$
which implies that $A_0(0) = \rho_e (n+1)/e^2v^2$. From Eq.(3.1), we
see that $A_0 \sim (n+1 - eA_\varphi)$.

Since we are interested in regular  solutions, there is one
parameter near $s=0$ is to  be adjusted to get the right asymptotic
behavior at $r=\infty$.
$$  f /v =   a s^n ( 1 - {1\over 4}s^2 +... )
\eqno\eq  $$
where the dots indicate the even power series in $s$ with higher power than
what is shown and
whose coefficients are fixed in terms of $a$ and $b$.
For large $r$, the regular solution would be
$$  f /v =  1- {c \over \sqrt{s}} e^{-s}
\eqno\eq $$
in leading orders.

As Eq.(3.2) is identical to the selfdual equations without any
background electric charge density, their properties have been well
studied numerically. The new aspect here is that the behavior of the
electric charge density, $e^2f^2 A_0 - \rho_e $, is nontrivial.
 At the origin the
Higgs field vanishes and so the background electric charge is exposed.
As this exposed charge is screened, the total electric charge density
would go from negative to positive and then falls to zero
exponentially.

\FIG\figA{ The vortex configuration of unit vorticity}
\FIG\figB{ The vortex configuration of $n=2$}

Fig.~1 shows the field configuration of a unit vorticity after scaled
fields in the distance scale $evr$. It shows the scalar field $f/v$,
the magnetic field $F_{12}/2ev^2$, the scalar potential $A_0/(\rho_e
/e^2v^2) $, and the electric charge density $evr[ e^2f^2 A_0 - \rho_e]
/\rho_e $. Fig.~2  describes the  field configurations for $n=2$.

\chapter{Linear and angular momenta}

 Our vortices carry nonzero magnetic flux but no net electric charge.
Thus, we can consider our system on a large sphere rather than on a
plane.  On the sphere there would be a conserved angular momentum
vector, which remains finite in the infinite volume limit.  In this
limit, the angular momentum vector would become the linear and angular
momenta on the plane. Here we will simply guess the correct answer on
the plane. We first notice that when there are vortices, the net
magnetic flux on sphere is nonzero and that the vector potential $A_i$
would have a Dirac string. The vector field
$$ \bar{A}_i \equiv \partial_i \theta + eA_i
\eqno\eq $$
is gauge invariant and well-defined except at the position of vortices
where $f=0$.  $\bar{A}_i$ would also carry the same magnetic flux as
$A_i$. Thus there should be Dirac strings for $\bar{A}_i$, whose only
possible positions are at the center of  vortices where $\bar{A}_i$ is
ill-defined. We want express the momenta in terms of $\bar{A}_i$
rather than $A_i$.

First, we use the field equations to  rewrite the
energy momentum tensor (2.3) as
$$ T_{\mu\nu} =  T^S_{\mu\nu} + \partial^\rho (F_{\rho\mu} A_\nu)
+ \rho_e (\eta_{\mu\nu} A_0 - \eta_{\mu 0}A_\nu )
\eqno\eq $$
where the nonconserved symmetric energy momentum tensor is defined by
$$ T^S_{\mu\nu} = - F_{\mu\rho}F_\nu^{\,\,\,\rho} + \partial_\mu N
\partial_\nu N + \partial_\mu f \partial_\nu f
+ f^2 (\partial_\mu \theta + eA_\mu ) (\partial_\nu \theta + eA_\nu)
- \eta_{\mu\nu} \left( {\cal L} + \rho_e A_0 \right)
\eqno\eq $$
Note that $T_{00} = T^S_{00} $ to a total space derivative.
The angular momentum current (2.7) can be rewritten as
$$ {\cal J}_\mu = -\epsilon_{ij} x^i T_{\mu j}^S
-\epsilon_{ij} \partial^\rho (x^i F_{\rho\mu} A_j)
- \rho_e \epsilon_{ij} x^i( \eta_{\mu j} A_0 - \eta_{\mu 0} A_j)
\eqno\eq $$
After throwing away any total derivatives, we still get a divergent
contributions to the angular momentum $\tilde{J} = \int d^2 x {\cal
J}_0 $ for the vortex configurations.

Now  consider a generic $\theta$ configuration given as
$$ \theta = - \sum_a (-1)_a {\rm Arg} (\vec{r} -\vec{q}_a(t))
+ \eta \eqno\eq $$
where $\eta$ is  a single-valued function. We then introduce a current
$$ {\cal I}_\mu = -{\rho_e \over e} \epsilon_{ij} x^i (\eta_{\mu j}
\dot{\theta} - \eta_{\mu 0} \partial_j \theta )
\eqno\eq $$
whose divergence is
$$ \eqalign{ \partial^\mu {\cal I}_\mu &\
= -{\rho_e \over e} \epsilon_{ij} x^i (\partial_j \partial_0 - \partial_0
\partial_j)
\theta \cr
&\ = {\pi  \rho_e \over e}
\sum (-1)_a {d |\vec{q}_a|^2 \over dt} \delta(\vec{r} -\vec{q}_a)
\cr}
\eqno\eq $$
If we neglect the divergence from the spatial infinity and
the mild singularity at the origin, we get a conserved charge,
$$ \Delta J = \int d^2 x {\cal I}_0 - {\pi \rho_e \over e} \sum_a
(-1)_a |\vec{q}_a|^2
\eqno\eq $$

We introduce  the new  angular momentum, as the sum of
$\tilde{J} + \Delta J$,
$$ \eqalign{ J   =   - \int d^2 x
\epsilon_{ij} x^i &\ \left\{   F_{0k}F_{jk} + \dot{N} \partial_j N
+\dot{f} \partial_j f + f^2 (\dot{\theta} +eA_0)
(\partial_j  \theta + eA_j) \right. \cr
&\  \left.  - {\rho_e \over e }
(\partial_j \theta + eA_j) \right\}
 - {\pi \rho_e \over e} \sum_a (-1)_a |\vec{q}_a|^2
\cr}
\eqno\eq  $$
The  divergent contributions at the spatial infinity cancel each
other in $J$ since $\partial_j \theta + eA_j$ falls off exponentially.
There are also no divergent contributions from the
vortex positions
because $\partial_i \theta \sim 1/|\vec{x} - \vec{q}_a|$. This  angular
momentum is also conserved. We also note that $J$ would generate the
same transformation of the fields as the canonical angular momentum
$\tilde{J}$. Hence the angular momentum given by Eq.(4.9) is as good
as we hope for. The well-defined linear momentum is obtained from the simple
observation that under the spatial translation, $\vec{x} \rightarrow
\vec{x} + \vec{a}$, $J \rightarrow J + \epsilon_{ij} a^i P^j$. This
linear  momentum is
$$ \eqalign{ P^i = - \int d^2x &\ \left\{ F_{0k}F_{ik} + \dot{N}
\partial_i N + \dot{f}\partial_i f + f^2(\partial_0 \theta + eA_0)
(\partial_i
\theta + eA_i) \right. \cr
&\ \left. - {\rho_e \over e} (\partial_i + eA_i) \right\} +
  {2\pi \rho_e \over e}  \epsilon_{ij} \sum_a (-1)_a  q_a^j
\cr}
\eqno\eq $$

We are now in the position to explore  the angular momentum for the selfdual
vortex
configurations. These configurations are time-independent and satisfy
Eqs.(2.17) and (2.13).  Since there are no divergent contributions
from the vortex position, we can take out these positions from the
integration domain without changing the value of the angular momentum.
We call this domain of integration as  $R^2_* \equiv  R^2 -
\{\vec{q}_a\} $. Eq.(4.9) for these selfdual configuration becomes
$$\eqalign{ J + {\pi \rho_e \over e} \sum_a |\vec{q}_a|^2 &\ = -
{1\over e} \int_{R^2_*} d^2x \left\{ x^i \partial_i A_0 \bar{F}_{12} +
\epsilon_{ij} x^i \partial_k^2 A_0   \bar{A}_j \right\} \cr
&\ = -{1\over e} \int_{R^2_*} d^2x \partial_i
\left\{ \epsilon_{jk} x^j \left( \partial_i A_0 \bar{A}_k - A_0
\partial_k \bar{A}_i \right) - \epsilon_{ij} A_0 \bar{A}_j \right\}
\cr
&\ =  {1\over e} \sum_a
\oint dl_a^i \left\{ \epsilon_{jk} (q_a^j + l_a^j ) \left(
\partial_i A_0 \bar{A}_k - A_0 \partial_k \bar{A}_k \right)
-\epsilon_{ij} A_0 \bar{A}_j \right\} \cr}
\eqno\eq $$
where the line integration is over the vortex positions and $l_a^i =
x^i - q_a^i $. The above angular momentum gets  separated naturally
into two pieces:  the extrinsic part proportional to  $q_a^i$
and the intrinsic part proportional to $l_a^i$.

\def\bA{ \bar{A}}

To evaluate the total angular momentum, let us study the behavior of
the fields near $\vec{q}_a$.
As $f \sim |\vec{r}-\vec{q}_a|$ at each vortex, we can expand $f$ near
$\vec{q}_a$ to
get
$$ \ln f = \ln |\vec{l}_a| + a_a  + b^i_a l^i_a + c^{ij}_a l^i_a
l^j_a + {\cal O}(l_a^3)  \eqno\eq $$
with $\vec{l} = \vec{r} -\vec{q}_a$.
Eq.(2.20)  implies then
$$ \sum_i c^{ii}_a = -{1\over 4} e^2v^2  \eqno\eq $$
Eqs.(4.1) and (2.17)  then  imply
$$ \bA_i  = \epsilon_{ij} \left[ {l_a^j \over |\vec{l}_a|^2 }
+ b^j_a + 2c_a^{jk}l_a^k \right] + {\cal O }(l_a^2) \eqno\eq
$$
We also get from Eq.(2.21)
$$ \partial_i A_0 = {\rho_e \over e^2v^2} \left\{ \left[ 1 - \sum_b
(\vec{q}_a-\vec{q}_b)\cdot{\partial \over \partial \vec{q}_b} \right]
b^i_a  + 4c^{ij}_a l_a^j \right\} + {\cal O} (|\vec{l}_a|^2)
\eqno\eq $$

We calculate the line integrals in Eq.(4.11) by using $\oint dl_a^i l_a^j /
|\vec{l}_a|^2 = \pi \delta_{ij}$.
By using Eq.(4.14), we get the intrinsic part as
$$ \eqalign{J_{int} &\  = {1\over e} \sum_a \oint dl_a^i
\left\{ \epsilon_{jk}l_a^j \left( A_0 \partial_k \bar{A}_i -
\partial_i A_0\bar{A}_k \right)  + \epsilon_{ij} A_0 \bar{A}_j
\right\} \cr
&\ =  0 \cr
}
\eqno\eq $$
The extrinsic part with Eq.(4.14) becomes
$$\eqalign{ J_{ext} &\ =- {1\over e} \oint dl_a^i \epsilon_{jk} q_a^j
\left(  A_0 \partial_k \bar{A}_i - \partial_i A_0 \bar{A}_k \right) \cr
&\ =
- {1\over e} \sum_a \oint dl_a^i \epsilon_{jk} q_a^j \left\{
(A_0(\vec{q}_a) + l_a^l \partial_l A_0 ) \epsilon_{in} \left[ {\delta^{nk}
\over |l_a|^2 } - {2 l^k_a l^n_a \over |l_a|^4} \right] -
 \partial_i A_0  \epsilon_{kn}{  l^n \over |l_a|^2} \right\}
\cr
&\ = - {2\pi \over e} \sum_a q_a^i \partial_i A_0 (\vec{r} =
\vec{q}_a) \cr}
\eqno\eq
$$
With  Eq.(4.15), the total angular momentum (4.11) can be expressed as
$$ J = - {2\pi \rho_e \over e^3v^2} \sum_a \vec{q}_a \cdot \left( 1 - \sum_b (
\vec{q}_a -\vec{q}_b)\cdot{\partial \over \partial \vec{q}_b} \right)
 \cdot \vec{b}_a -{\pi \rho_e \over e} \sum_a |\vec{q}_a|^2
\eqno\eq $$
The angular momentum is now expressed rather explicitly in terms of
the selfdual configurations.

Since the scalar potential $A_0$ is a smooth function of $\vec{r}$ and
$\vec{q}_a$, the external part (4.17) is a smooth function of the
vortex position. When all vortices come together at the origin, the
configuration is symmetric and so the total angular momentum vanishes.
This means that these vortices do not carry any spin, contrast to the
claims in Ref.[2]  The part of the angular momentum which is proportional
to $\sum_a |\vec{q}_a|^2$ is due to the Magnus force by the shielding
charge. Such a term arises whenever charged particles move in a uniform
magnetic field. We will derive Eq.(4.18)  in the next section via a rather
different path.

If we have considered a general Maxwell-Higgs system, vortices would
have a static force between them and would not stay stationary in
general. We can still calculate the spin of vortices by considering a
single vortex sitting at the origin. Following the similar line of
reasoning as before and using Eq.(4.9) which is correct for any
potential, one can see easily that the spin of vortices is zero.

For two vortices located at points $\vec{q}_1 =\vec{q}/2$ and
$\vec{q}_2= -\vec{q}/2$, the symmetry of the configuration implies
that $f(\vec{r};\vec{q}) = f(\vec{r};-\vec{q}) =
f(-\vec{r};-\vec{q})$,
which in turn implies $\vec{b}_1(\vec{q}) = -\vec{b}_1(-\vec{q}) =
-\vec{b}_2(\vec{q})$. In addition, the $f$ configuration is
invarariant under the reflection which exchanges two vortices,
implying
$$ \vec{b}_1 = \hat{q}{\cal B}(q)   \eqno\eq $$
where $q = |\vec{q}|$. The conserved angular momentum (4.18) becomes
$$ J = - {\pi \rho_e \over 2e}q^2 -{2\pi \rho_e \over e^3v^2} q (1+ q{
d \over dq}) {\cal B}(q)
\eqno\eq $$
We show in the appendix that ${\cal B}(q) = 1/q  + {\cal
O}(q)$ near $q=0$. Thus Eq.(4.20) goes to zero when $q=0$, which again
implies  our vortices carry no spin.

A rather similar expression as in Eqs.(4.18) and (4.20) has been
obtained for vortices in the selfdual Chern-Simons Higgs
systems.$^{[7]}$ In that case, vortices carry nonzero spin $s_v$ and the
total angular momentum of two vortices decreases from $4s_v$ to $2
s_v$ as the distance between two vortices increases from zero to
infinity.

\chapter{Slow motion of vortices}

The selfdual configurations of $n$ vortices are characterized by the
vortex positions, $\vec{q}_a$. Let us now consider the dynamics of
slowly moving vortices. When they move slowly enough, we expect that
the classical radiation is very small and their dynamics is described
by the  Newtonian mechanics.  Thus, we hope that their dynamics could be
summerized by an effective Newtonian lagrangian or action. Our goal
would be to find this effective action, which would be expressed in
terms of the selfdual configurations.

We expect that the field configuration for a given trajectory
$\{\vec{q}_a(t) \}$ of slowly moving vortices is very close to the
selfdual configuration because there is no or little radiation. Even
though the Gallileo transformation is no longer a symmetry of the
system due to the background electric charge density, we can get a
hint for the field configuration of a slowly moving vortices from this
transformation. When there is no background charge density, the
configuration for uniformly moving vortices is corrected linearly in
velocities  (especially the vector field) and satisfy the lagrangian
field equation to the first order in velocities.$^{[7]}$

Thus, it seems sensible to assume that the field configuration of
slowly moving vortices has first order corrections in general, as
$f(\vec{r},\vec{q}_a(t)) +
\Delta f$, $N(\vec{r},\vec{q}(t)) + \Delta N$,
$A_\mu(\vec{r},\vec{q}_a(t))+
\Delta A_\mu$, and $\theta = - \sum_a {\rm Arg } (\vec{r} -\vec{q}_a)$.
and that they satisfy the field equations to the first order in
velocity. We have chosen a gauge where there is no correction to
$\theta$. The zeroth order terms satisfy the selfdual equations,
(2.12) and (2.13) with the upper sign. The first order corrections
then satisfy the the field equations to the linear order in
velocities,
$$ \eqalign{ &\ \partial_i^2 \Delta f + e^2 A_0^2 \Delta f -
(\partial_i \theta + eA_i)^2 \Delta f -e^2N^2 \Delta f - {e^2 \over 2}
(3f^2 - v^2) \Delta f \cr
&\ \,\,\,\,\,\,\,\,\,\,\,\,\,\,\,\,\,\,
 +2efA_0 (\dot{\theta} +
e\Delta A_0) -2e f(\partial_i \theta + eA_i) \Delta A_i- 2e^2fN\Delta
N = 0 \cr
 &\  \partial_i \dot{A}_i - \partial_i^2 \Delta A_0 +
ef^2(\dot{\theta} +e \Delta A_0) + 2e^2A_0 f \Delta f = 0 \cr
&\  \partial_i \dot{A}_0 - \epsilon_{ij}\partial_j \Delta F_{12}-
e^2 f^2 \Delta A_i - 2e(\partial_i \theta + eA_i) f \Delta f = 0 \cr
&\ \partial_i^2 \Delta
N - e^2 f^2 \Delta N -2e^2 Nf \Delta f = 0 \cr }
\eqno\eq $$

We evaluate the origin field action for this corrected field
configurations for slowly moving vortices. The zeroth order lagrangian
will be simply the negative of the total mass. The first order correction
is given by
$$ \eqalign{ \Delta_1 {\cal L} = &\
(\dot{A}_i - \partial_i \Delta A_0 )(-\partial_i A_0) - F_{12} \Delta F_{12}
- \partial_i N\partial_i \Delta N - \partial_i f \partial_i \Delta f \cr
&\ + e^2 A_0^2 f\Delta f - (\partial_i \theta + eA_i)^2 f \Delta f
 + ef^2A_0(\dot{\theta} + e\Delta A_0) - ef^2 (\partial_i \theta + eA_i)\Delta
A_i
\cr
&\ - {e^2 \over 2}(3f^2 -v^2) \Delta f - e^2 N^2 f \Delta f
- e^2 f^2 N \Delta N + \rho_e (\Delta N - \Delta A_0) \cr
} \eqno\eq $$
Since the zeroth field configurations  satisfy  the time independent
 field equations, the above expression can be simplified as
$$ \Delta_1 {\cal L} = - \partial_i A_0 \dot{A}_i + e f^2 A_0 \dot{\theta}
 \eqno\eq $$
The selfdual equations (2.17) imply  that the zeroth order $A_i$ is
transverse, making the first term of $\Delta_1 L$
a total derivative.   We now use Gauss's law (2.13) to get
$$ \eqalign{   \Delta_1  L[q_a^i,\dot{q}_a^j]
 &\   = {1\over e} \int d^2x  (\rho_e +
\partial_k^2 A_0) \dot{\theta}  \cr
&\ = -{1\over e} \sum_a \epsilon_{ij} \dot{q}_a^i \int d^2x
(\rho_e + \partial_k^2 A_0 ) \partial_j \ln |\vec{r} - \vec{q}_a| \cr
}  \eqno\eq $$
where $\dot{\theta} = - \sum_a \epsilon_{ij} \dot{q}_a^i
\partial_j \ln |\vec{r}-\vec{q}_a|$ is used.

The right hand side of Eq.(5.4) is infrared divergent.
To understand the infrared divergent term, let us define
$$ V^i_a (\vec{q}_b) = -{\epsilon_{ij}\over e}
 \int d^2_{R^2} x (\rho_e  + \partial_l^2 A_0)
 \partial_j \ln | \vec{r}  -\vec{q}_a |  \eqno\eq $$
Since the contribution from the point $\vec{r} = \vec{q}_a$ is
nonsingular, we take out this point from the integration domain.
$V_a^i$ is a vector potential as  $\Delta_1 L = \sum_a \dot{q}_a^i
V_a^i $.  Since   $(\rho_e +
\partial_k^2 A_0)(\vec{r} = \vec{q}_a) = 0 $, the
curl of $V_a^i $  is
$$ \eqalign{ {\partial V_a^i \over \partial q_b^k } &\
= - {\epsilon_{ij} \over e}
 \int_{R^2 -\vec{q}_a} d^2  x
\left\{ \left[ {\partial \over
\partial q_b^l} \partial_l^2 A_0 \right]  \partial_j \ln |\vec{l}_a|
 + (\rho_e + \partial_l^2 A_0) (-\delta_{ab}) \partial_k
\partial_j \ln| \vec{l}_a | \right\}   \cr
&\  = - {\epsilon_{ij}\over e}
\int_{R^2-\vec{q}_a} d^2x
 \partial_l \left\{ \left[ \partial_l
 {\partial A_0 \over \partial
q_b^k} \partial_j \ln |\vec{l}_a | -  {\partial A_0 \over \partial
q_b^k }  \partial_l \partial_j \ln |\vec{l}_a| \right]
  -\delta_{ab} \left[ \partial_l A_0 \partial_k \partial_j \ln
|\vec{l}_a| - A_0 \partial_l \partial_k \partial_j \ln |\vec{l}_a|
\right] \right\} \cr
&\ = {\epsilon_{ij}\over e}
\oint dl_a^l \left\{ \biggl[\partial_l {\partial A_0 \over \partial q_b^k}
\partial_j \ln |\vec{l}_a|  -  {\partial A_0 \over \partial q_b^k }
\partial_l \partial_j \ln |\vec{l}_a|  \biggr] -
\delta_{ab} \biggl[\partial_l A_0 \partial_k \partial_j \ln|\vec{l}_a|
-A_0 \partial_l\partial_k\partial_j \ln |\vec{l}_a| \biggr] \right\}
\cr
&\ = \left. {2\pi \epsilon_{ij} \over e}
\left\{   \partial_j {\partial A_0
\over \partial q_b^k}
- \delta_{ab}  \left[ {1\over 2} \delta_{jk}
 \partial_l^2 A_0 - \partial_j \partial_k A_0 \right]\right\}
\right|_{\vec{r} = \vec{q}_a}
 \cr}
\eqno\eq $$
where we used  $\oint dl^i l^j /|\vec{l}^2| = \pi \delta^{ij} $ and
its generalizations.
We use Eqs. (4.13) and  (4.15) to get
$$  {\partial V_a^i \over \partial q_b^k }
= {2\pi \rho_e \epsilon_{ij} \over e^3 v^2}
\left\{ { e^2v^2 \over 2} \delta_{ab} \delta_{jk}
  + [ 2 - \sum_c(\vec{q}_a - \vec{q}_c)\cdot{\partial
\over \partial \vec{q}_c} ]  {\partial b_a^j \over \partial q_b^k }
\right\}
\eqno\eq $$
which indicates that the linearly divergent part is independent of the
$\vec{q}_b$.  we have used here the relation $\sum_c \partial b_a^i /
\partial q_c^k = 0$ due to the translation invariance.
 Integrating Eq.(5.7), we can choose a gauge so that
$$ V_a^i =    {2 \pi \rho_e \epsilon_{ij} \over e^3 v^2}
\left\{  {e^2v^2 \over 2} q_a^j + [ 1 -\sum_b (\vec{q}_a - \vec{q}_b) \cdot
{\partial \over \partial \vec{q}_b} ] b_a^j \right\}
\eqno\eq $$

Even though $b_a^i$s fall to zero exponentially when the mutual
distances between vortices increase, it does not mean there is no
nontrivial statistical phase when two vortices are exchanges because
the above derivation assumes that vortices are not overlapped and so
could miss a singular potential with zero magnetic field when vortices
are separated. As we will see later, a better guide would be whether a
vortex feels any additional magnetic field besides the average uniform
magnetic field due to the presence of other vortices.

The second order correction to the effective lagrangian
is given by
$$ \eqalign{ \Delta_2 {\cal L} =
&\ {1\over 2} (\dot{A}_i - \partial_i \Delta A_0)^2 - \partial_0
\Delta A_i \partial_i A_0  - {1\over 2} (\Delta F_{12})^2
+ {1\over 2} [ \dot{N}^2 - (\partial_i \Delta N)^2 ] \cr
&\ {1\over 2} [\dot{f}^2 - (\partial_i \Delta f)^2 ]
+ {1\over 2} f^2 [ (\dot{\theta} +e\Delta A_0)^2 - e^2 (\Delta A_i)^2]
\cr
&\ + {1\over 2} [\Delta f)^2 ( e^2A_0^2 - (\partial_i \theta + eA_i)^2]
+ 2f\Delta f [ e A_0 (\dot{\theta} + e\Delta A_0) - e(\partial_i
\theta + eA_i)\Delta A_i ] \cr
&\ -{e^2 \over 4} (3f^2 -v^2) (\Delta f)^2 - e^2 N^2 (\Delta f)^2
-e^2 f^2 (\Delta N)^2 - 2e^2 f\Delta f N \Delta N \cr
}\eqno\eq $$
We can imagine the possible contribution from  the second
order corrections of  the field configuration to $\Delta_2 {\cal L}$.
However, the first order field equation implies this possible contribution
vanishing, making our approximation consistent.
After using Eq.(5.1) satisfied by the first order corrections,
we get
$$ \Delta_2 L[q_a^i,\dot{q}_a^j]
 = \int d^2x \left\{ {1\over 2} \bigl[\dot{A}_i^2 + \dot{N}^2
+ \dot{f}^2 \bigr] + {1\over 2} \bigl[\Delta A_i \partial_i \dot{A}_0
+ { 1\over e} \Delta A_0 \partial_i^2 \dot{\theta} \bigr] \right\}
\eqno\eq $$
In principle, we can solve Eq.(5.1) and express the first order
corrections of the fields in terms of the selfdual configurations,
leading to $\Delta_2 L$ fully expressed in the selfdual
configurations. Samols in Ref.[9]  managed to express the second order
terms explicitly for the case when there is no background electric
charge density. It would be interesting to find such expression in our
case too.

We have now an effective lagrangian for slowly moving vortices.  Since
the energy of vortices at rest does not depend on the vortex
positions, there is no static force between them. However, the
shieding charge carried by the Higgs field manifest itself as a
uniform magnetic field acting on vortices. The Magnus force due to the
shielding charge density is now a Lorentz force by this magnetic
field.  In addition, there is a magnetic interaction between vortices
because the shielding charge density around vortices is not uniform.

Let us study the effective action for slowly moving vortices in more
detail.  The effective lagrangian can be written figuratively as
$$ L_{eff}[q_a^i,\dot{q}_a^j]  = {1\over 2} \sum_{ab,ij} T_{ab}^{ij}(\vec{q}_c)
\dot{q}_a^i \dot{q}_b^j + \sum_{a,i} \epsilon_{ij} \dot{q}_a^i
\bigr[ \alpha q_a^j + H_a^j(\vec{q}_c) \bigl]
\eqno\eq $$
where
$$ \eqalign{ &\ \alpha = {\pi \rho_e \over e} \cr
&\ H_a^i = {2\pi \rho_e \over e^3v^2} \left\{ 1 - \sum_b (\vec{q}_a
-\vec{q}_b)\cdot {\partial \over \partial \vec{q}_b} \right\} b^i_a \cr
}\eqno\eq $$
The dynamical equation for the a-th vortex is
$$\eqalign{  { d (T_{ab}^{ij} \dot{q}_b^j) \over dt } - {1\over 2}
{\partial T_{bc}^{jk} \over \partial q_a^i } \dot{q}_b^j \dot{q}_c^k
&\ = -2\alpha \epsilon_{ij} \dot{q}^j_a
 + \left\{ \epsilon_{jk} {\partial H_b^k \over \partial q_a^i } -
\epsilon_{ik} {\partial H^k_a \over \partial q_b^j} \right\}
\dot{q}_b^j
\cr
&\ = -2\alpha \epsilon_{ij} \dot{q}_a^j -\epsilon_{ij} \dot{q}_b^j
{\partial H_b^k \over \partial q_a^k} +
\epsilon_{ij} \dot{q}^j_b \left( {\partial H_b^k \over \partial q_a^j }
- {\partial H_a^k \over \partial q_b^j } \right) \cr}
\eqno\eq $$
For a given positive $\rho_e$,  the parameter
$\alpha$ are positive. The Magnus force due to the shielding charge
from the above equation of motion makes  a
moving vortex  turn left.

The effective lagrangian (5.11) is invariant under time translation and
so the tensors $T , H$ are independent of time. The conserved energy
is then
$$ E_{eff} = {1\over 2} \sum_{ab,ij}   T_{ab}^{ij} \dot{q}_a^i \dot{q}_b^j
\eqno\eq $$
The space translation invariance implies that the tensors $T, H$ are
invariant under $\vec{q}_a \rightarrow \vec{q}_a + \vec{\epsilon} $, or
$$ \sum_c {\partial T_{ab}^{jk}  \over \partial q_c^i } = 0
= \sum_c {\partial H_a^j \over
\partial q_a^i }
\eqno\eq $$
The conserved linear momentum is then
$$ P_{eff}^i = \sum_a \left\{   T_{ab}^{ij} \dot{q}_b^j +
\epsilon_{ij}[ 2\alpha q_a^j  +H_a^j] \right\}
\eqno\eq $$
The rotational invariance implies that
$$ \eqalign{ &\
\epsilon_{ij} H_a^j + \epsilon_{kl} q_b^k {\partial H_a^i \over
\partial q_b^l} = 0 \cr
&\ \epsilon_{ik} T_{ab}^{kj} + \epsilon_{jk} T_{ab}^{ik}
+ \epsilon_{kl} q_c^k {\partial T_{ab}^{ij} \over \partial q_c^l} = 0 \cr
} \eqno\eq $$
The conserved angular momentum is given as
$$ J_{eff} = \epsilon_{ij} q^i_a T_{ab}^{jk} \dot{q}_b^k - \alpha |
\vec{q}_a|^2 - q_a^i H_a^i
\eqno\eq $$
Under the translation $\vec{x} \rightarrow \vec{x} + \vec{a}$, the
angular momentum transforms as $J \rightarrow J + \epsilon_{ij} a^i P^j$.

We can calculate the energy, linear momentum and angular momentum for
our configurations of slowly moving vortices from Eqs.(2.4), (4.10) and
(4.9).  The interesting question is whether for our slowly moving
vortices these field theoretical conserved quantities are  identical
to those from the above effective action. For vortices at rest, we can
compare easily. The total energy in the field theory would be just the
total mass. The kinetic energy vanishes.  The total angular momentum (5.18)
from  the effective
action with Eq.(5.12) becomes
$$ J_{eff} = - {\pi \rho_e \over e} \sum_a |\vec{q}_a|^2
 - { 2\pi \rho_e \over e^3v^2}   \sum_a \vec{q}_a \cdot
\left\{ 1 - \sum_b(\vec{q}_a - \vec{q}_b){\partial \over \partial \vec{q}_b}
\right\} \vec{b}_a   + {\cal O}(\dot{q})
\eqno\eq
$$
This is exactly what is given in Eq.(4.18). Since these results are
agreeing each other, we have some confidence  in the linear part of our
effective action.

We have the effective lagrangian for slowly moving vortices. Let us
first apply the effective action to a single vortex at the position
$\vec{q}$. The effective action would be then just
$$ L = {1\over 2} m\dot{q}^2 + {\pi \rho_e \over e} \epsilon_{ij}
\dot{q}^i q^j
\eqno\eq $$
where $T_{aa}^{ij} = m \delta^{ij}$ with the particle mass $m$.  (We
have not shown  that this mass is the rest mass of
vortices.) This is a lagrangian for a charged particle moving on a
uniform magnetic field background. The Magnus force manifests itself
as the Lorentz force due to this magnetic field. Since our vortices do
not carry any spin and the Magnus force is usually associated with
nonzero spin, we are in a somewhat ironic situation. A single vortex would
move a circle. Quantum mechanically there will be a nonzero Berry's
phase on the wave function when the vortex moves around a closed loop.
The argument of the phase would be proportional to the total magnetic
flux encircled by the loop,
$$  {\rm phase}  = \exp \left\{ i{\pi \rho_e \over e} \oint_C \epsilon_{ij}
dq^i
q^j \right\}
=  \exp \left\{ i {\pi \rho_e \over e} {\rm area} \right\}
\eqno\eq $$

Let us now
consider the system of two vortices moving with the positions
at $\vec{q}_1 = \vec{q}/2,
\vec{q}_2 = -\vec{q}/2$ as before. The total angular is now given in
Eq.(4.20). The first order part of the effective action from
Eqs.(5.11) and Eq.(5.12) becomes
$$ \Delta_1 L = \epsilon_{ij}  \dot{q}^i q^j \left\{
{\pi\rho_e \over 2 e}  + {\pi \rho_e \over e^3v^2}  {d q{\cal B}(q) \over dq}
\right\}
\eqno\eq $$
The magnetic field felt by the reduced system is given by
$$ {\cal H}_{12}(q) = {d J_{eff} \over qdq} \eqno\eq $$
Since $q {\cal B}= 1$ at $q=0$ and ${\cal B}$ goes zero exponentially at the
spatial infinity, one can deduce from Eqs.(4.20) and (5.22) that
the total magnetic flux felt by the reduced particle
when it goes around a circle of large $q$ would be
just the total area times the flux $\rho_e /e$.
This implies that there is no nontrivial statistical phase between two
vortices in large separation, proving the spin-statistics theorem.
In finite distance, the matter is more complicated. Obviously ${\cal
B}(q)$ is a complicated function of $q$ and would lead to an
interesting dynamics of two vortices.

By using the spin-statistcs theorem, we can argue that our vortices
should not carry any spin.  The spin-statistcs theorem in three
dimensions implies that particles and antiparticles carry the same
sign spin $s$. When there is no background magnetic field, the
statistics works out because the theorem implies that the orbital
angular momentum between two partices is $2l + 2s$ and that between a
particle and an antiparticle is $2l -2s$. In our case vortices and
antivortices have the same charge profile and the opposite electric
current. Thus, if they carry any spin, the spin of vortices should
have the opposite sign to that of antivortices. Since CTP is a good
symmetry of the theory, the spin-statistics theorem should however be
correct, implying the same sign. Hence there is no conflict if
vortices do not carry any spin, which we have shown.

\chapter{Conclusion}

We has studied the vortex dynamics in selfdual Maxwell-Higgs systems with
uniform background electric charge density. We have found a
well-defined  modification of  the Noether angular
momentum.  Our vortices are shown to carry no spin
but feels the Magnus force due to the shielding charge carried by the
Higgs field.  We have studied dynamics of
the slowly moving vortices, proving the spin-statistics theorem of
vortices. There are many directions we can take from here. The further
investigation of the slowly moving vortices, especially quadratic
terms would be interesting.

\centerline{\bf Acknowledgement}

The author thank Choonkyu Lee, Mel Ruderman, Erick Weinberg for useful
discussions. He  also thank the organizers of the
small scale structure of spacetime workshop in ITP where the part of
this work is done.

\appendix

Here we calculate the contribution of the symmetric energy tensor to the
angular momentum of the selfdual vortex configurations. The angular
momentum is
$$J^S = - \int d^2x \epsilon_{ij} x^i T_{0j}^S
\eqno\eq $$
where the symmetric energy momentum tensor is given in Eq.
This is not conserved but well defined without any divergent
contribution from the spatial infinity or the vortex positions. It
would become a useful quantity to characterize the static configurations.
For the selfdual configurations satisfying Eqs.(2.17) and (2.13)   $J^S$
becomes
$$ \eqalign{  J^S  &\   = - \int d^2 x \left\{x^i \partial_i A_0 F_{12} +
e f^2 A_0 \epsilon_{ij}x^i (\partial_j \theta + eA_j) \right\} \cr
&\  =   {e\over 2}  \int d^2x x^i\partial_i[A_0(f^2-v^2)] \cr
&\ =    e \int d^2r A_0(v^2 - f^2) \cr
&\ =    {\rho_e \over e} \int d^2r [{ e^2v^2 A_0 \over \rho_e}  - 1]
 \cr}
\eqno\eq $$
Antivortices would have the same charge profile but the opposite
current. Thus $J^S$ of antivortices would have the opposite sign to
that of vortices.

We can express $J^S$ more explicitly for the selfdual configurations.
To achieve this goal, first we express the angular momentum in terms
of $\bar{A}_i \equiv
\partial_i \theta + e A_i $. As there is no singular contribution at
vortex positions $\{q_a \}$ in Eq.(A.2), we take out these points from
the integration domain. From Eqs. (2.21) and (2.17),  we get
$$ \partial_i A_0 = - {\rho_e \over e^2 v^2}
\epsilon_{ij} (1 + {\cal D})  \bar{A}_j
\eqno\eq $$
where
$$ {\cal D} \equiv -\sum_a (\vec{r} - \vec{q}_a)\cdot {\partial \over
\partial \vec{q}_a }  \eqno\eq $$
{}From Eqs.(2.13) and (A.3),  we get
$$ \eqalign{ f^2 A_0 &\ = {\rho_e \over e^2 } + {1\over e^2}
\partial_i^2 A_0 \cr
&\  = {\rho_e \over e^2} - {\rho_e \over e^4v^2} [2 + {\cal D}] \bar{F}_{12}
\cr}
\eqno\eq $$
The nonconserved  angular momentum (A.2) becomes
$$ J^S = {\rho_e \over e^3 v^2 } \int_{R^2_*}  d^2x \left\{
- e^2 v^2 \epsilon_{ij} x^i \bar{A}_j +  \epsilon_{ij} x^i
(3+ {\cal D} ) [  \bar{A}_j \bar{F}_{12} ]
\right\}
\eqno\eq $$
{}From Eq.(2.17),  we observe that
$$ \eqalign{ -\epsilon_{ij} x^i \bar{A}_j ( e^2v^2  - 2 \bar{F}_{12} )
&\ =  - { e^2\over 2}  x^i \partial_i (1-f^2)  \cr
 &\ = e^2(v^2  - f^2)  \cr
&\  = 2\bar{F}_{12} \cr }
\eqno\eq $$
up to a total derivative.
With an identity $
\epsilon_{ij} \epsilon_{kl} = \delta_{ik} \delta_{jl} -
\delta_{il}\delta_{jk}$, $\partial_i \bar{A}_i = 0 $, and $  (1+
{\cal D}) \partial_i = \partial_i {\cal D} $, we get
$$ \eqalign{  J^S &\
 = {4\pi n \rho_e \over e^3v^2}   + {\rho_e \over e^3v^2}
\int_{R^2_*}  d^2 r \partial_i  {\cal D}\left[ x^j \left\{
{1\over 2} \delta_{ij}  \bar{A}_k^2
- \bar{A}_i \bar{A}_j \right\} \right]  \cr
&\ = {4 \pi n \rho_e \over e^3v^2} - \sum_a {\rho_e \over e^3v^2}
  \oint_{\vec{q}_a} dx^i x^j {\cal D} \left\{ {1\over 2}
\delta_{ij}  \bar{A}_k^2 -
\bar{A}_i \bar{A}_j \right\} \cr}
\eqno\eq $$
where each surface integration is done around the vortex positions $\{
\vec{q}_a \}$.

 When all vortices sit together at the origin,
$$ \ln f = n \ln |\vec{x}| + a + {\cal O}(x^2) \eqno\eq $$
due to the rotational symmetry. Thus,
$$\bar{A}_i = n \epsilon_{ij} {x^j \over |\vec{x}|^2} + {\cal O}(x)
\eqno\eq $$
Thus, the  angular momentum becomes
$$ J^S = {2\pi \rho_e \over e^3v^2} n(n+2) \eqno\eq $$
for the $n$ overlapped vortices.

For $n$ separated vortices, we separate the integration part of
Eq.(A.8)  to  two pieces,
by introducing $ \vec{l}_a = \vec{r} - \vec{q}_a$ with
$\oint_{\vec{q}_a } dx^i = \oint d l^i_a $.
First, let us calculate the intrinsic part,
$$ - \oint dl_a^i l_a^j {\cal D}[  {1\over 2} \delta_{ij}  (\bar{A}_k^2)
- \bar{A}_i \bar{A}_j ]  = 2\pi  \eqno\eq $$
with ${\cal D} = \vec{l}_a \cdot \partial/\partial \vec{r}
 - \sum_b (\vec{q}_a - \vec{q}_b) \cdot \partial/\partial \vec{q}_b
- \vec{l}_a \cdot (\sum_b   \partial/\partial \vec{q}_b +
\partial/\partial \vec{r}) $ and $\oint dl^i l^j /|\vec{l}|^2 = \pi
\delta_{ij}$.
Thus, Eq.(A.8) can be  written as
$$ J^S = {6\pi  \rho_e n\over e^3v^2}  -\sum_a {\rho_e \over e^3v^2}
\oint d l_a^i q_a^j {\cal
 D} \left\{ {1\over 2} \delta_{ij} \bar{A}_k^2 -\bar{A}_i \bar{A}_j
 \right\}
\eqno\eq $$
After some algebra by using Eqs.(4.13) and (A.14), we get
$$ J^S  = {6\pi \rho_e n\over e^3v^2}    + {2\pi \rho_e \over e^3v^2}
 \sum_a \vec{q}_a \cdot \left\{
1 + \sum_{b\neq a} (\vec{q}_a - \vec{q}_b)\cdot {\partial
\over \partial \vec{q}_b } \right\} \vec{b}_a  \eqno\eq $$
Note that $J^S(\vec{q})$ is a continuous function of the vortex
positions.

Consider two vortices at rest with positions $\vec{q}_1 = \vec{q}/2,
\vec{q}_2 = -\vec{q}/2$.  Following the argument before Eq.(4.19), we get
$$ J^S = {12\pi \rho_e \over e^3v^2} + {2\pi \rho_e \over e^3v^2}
q(1 - q{d \over dq}) {\cal B}(q)
\eqno\eq $$
which should approach $16\pi \rho_e /e^3v^2 $ when $q\rightarrow 0$
because of Eq.(A.11).
Thus,
$$ {\cal B} = {1\over q} + {\cal O}(1)
\eqno\eq $$
near $q=0$.

\endpage

\refout
\endpage
\figout
\end